\documentclass[10pt,epsfig,graphics,mathbbm]{article}

\usepackage{amsmath,amsfonts,amssymb,graphics,graphicx,epsfig,color,times,bbm}
\usepackage[sort&compress]{natbib}

\newcommand{\me}{\mathrm{e}}
\newcommand{\mi}{\mathrm{i}}
\newcommand{\md}{\mathrm{d}}

\newcommand{\cc}{\mathbbm{C}}
\newcommand{\zz}{\mathbbm{Z}}
\newcommand{\nn}{\mathbbm{N}}
\newcommand{\rr}{\mathbbm{R}}
\newcommand{\id}{\mathbbm{1}}
\renewcommand{\vec}[1]{\text{\boldmath$#1$}}

\newtheorem{theorem}{Theorem}
\newtheorem{lemma}{Lemma}

\newcommand{\proof}{{\em Proof. }}
\newcommand{\qed}{\hfill$\square$\par\vskip24pt}

\begin{document}

\title{\large\bf Locality of dynamics
in general harmonic quantum systems}

\author{\normalsize M.\ Cramer$^{1,2}$, A.\ Serafini$^{3}$, 
and J.\ Eisert$^{1,2}$}

\date{}
\maketitle

\vspace*{-.6cm}


\centerline{\it\footnotesize  
1 QOLS, Blackett Laboratory, 
Imperial College London,
Prince Consort Road, London SW7 2BW, UK}

\centerline{\it\footnotesize  
2 Institute for Mathematical Sciences, Imperial College London,
Prince's Gardens, London SW7 2PE, UK}

\centerline{\it\footnotesize  
3 Department of Physics and Astronomy, University
College
London, Gower Street, London WC1E 6BT, UK}

\begin{abstract}
The Lieb-Robinson theorem states that locality is approximately preserved
in the dynamics of 
quantum lattice systems. Whenever one has finite-dimensional
constituents, observables evolving in time
under a local Hamiltonian will essentially grow linearly in their
support, up to exponentially suppressed 
corrections. In this work, we formulate Lieb-Robinson bounds for
general harmonic systems on general lattices,
for which the constituents are infinite-dimensional, as systems
representing discrete versions of free fields or the harmonic
approximation to the Bose-Hubbard model. We consider
both local interactions as well as infinite-ranged interactions, showing
how corrections to locality are inherited from the locality of the 
Hamiltonian: Local interactions
result in stronger than exponentially suppressed corrections,
while non-local algebraic interactions result in algebraic suppression.
We derive bounds for canonical operators, Weyl operators and 
outline generalization to arbitrary operators.
As an example, we discuss
the Klein-Gordon field, and see how the approximate locality in
the lattice model becomes the exact causality in the field limit. 
We discuss the applicability of these results to quenched
lattice systems far from equilibrium, and the dynamics
of quantum phase transitions.
\end{abstract}

\section{Introduction}

Locality in relativistic theories ensures that space-like separated
observables commute: One simply
cannot communicate faster than light. In non-relativistic
lattice models, in contrast, there is no a priori reason for the support of 
time-evolved operators to stay confined within a light
cone. 
It is one of the classic
results of mathematical physics, 
dating back to Lieb and Robinson \cite{LiebRobinson},
that even in a non-relativistic quantum spin model on a lattice, locality
is preserved in an approximate sense: There is always
a well-defined velocity of information propagation and hence a causal, ``sound'' or ``light" cone.
 Locality is then respected 
under quantum spin dynamics with finite-ranged interactions,
in that the support of any local observable evolved
for some time will remain local to a region of size linear in this time, 
up to a correction that is at least exponentially suppressed. 
Except from exponentially decaying 
tails, hence, one encounters a situation very 
much like in the relativistic setting. The Lieb-Robinson theorem
has hence put the observation in many quantum lattice models
on a rigorous footing that there exists a well-defined finite
speed of propagation, often referred to as group
velocity. 

Over the years, Lieb-Robinson bounds have
been extended and generalized to higher-dimensional 
spin systems on lattices, and the bounds have been significantly
improved in several ways \cite{LiebRobinson,Hastings,Koma,Correlations,Sims,Schlein,New,Proc}.
Also, important new applications of the Lieb-Robinson theorem
have been found: Notably, the only known proof for the
clustering of correlations in gapped lattice models is based
on the Lieb-Robinson theorem \cite{Hastings,Correlations}, 
hence rigorously 
confirming the ``folk theorem'' in condensed matter physics
that correlation functions decay exponentially in gapped models.
``Area laws'' for the scaling of entropy in ground states of
one-dimensional spin systems
can be proven based on this result 
\cite{HastingsArea,Entanglement,Entanglement2}. 
Finally, in the context of quantum information theory, it 
provides a bound to the velocity one can transmit
quantum information through a chain of systems giving
rise to a quantum channel \cite{Entanglement2},
a topic that has received quite some attention in the 
quantum information literature.

In the simplest form of the Lieb-Robinson theorem, one considers a spin system 
${\cal H}=\bigoplus_{i\in L}{\cal H}_i$,
${\cal H}_i=\cc^d$, on a lattice
with vertices $L$, and a local (finite-ranged)
Hamiltonian
$\hat H=\sum_{X\subset L}
\hat \Phi(X)$. The
time evolution of an observable $\hat A$
on some subset $A\subset L$ of the lattice under this Hamiltonian,\footnote{This automorphism
group of time evolution is in the context of Lieb Robinson bounds
typically denoted as
\begin{equation}
	\tau_t^{\hat H}(\hat A)=\me^{\mi\hat Ht} \hat A \me^{-\mi \hat Ht}.
\end{equation}
Since in this work, the Hamiltonian $\hat H$ is always well-specified at the beginning
of each section, we make use of the above notation in the Heisenberg picture for simplicity
of notation.}
\begin{equation}
	\hat A(t)= \me^{\mi\hat Ht} \hat A \me^{-\mi \hat Ht},
\end{equation}	
then forms a group of automorphisms. At time zero, obviously
any observable $\hat A$ being supported on $A$ will commute with any
observable
$\hat B $ that is supported on a disjoint set $B\subset L$.
The Lieb-Robinson bound now gives a bound to this commutator if
$\hat A$ is evolved in time under this local Hamiltonian. It says that there
exists a constant $C>0$ and a ``speed of light"
$v>0$ such that 
\begin{equation}\label{so}
	||[\hat A(t),\hat B]||\leq C ||\hat A||\, ||\hat B||\,\me^{-\mu (dist(A,B)-v|t|)},
\end{equation}
where $dist(A,B)=\min_{a\in A, b\in B} dist(a,b)$ is the minimal distance between
the two 
regions and $\|\cdot\|$ denotes the operator norm. In other words, outside the causal cone $v|t|<dist(A,B)$, one encounters merely
an exponential tail, and the supports of $\hat A(t)$ and $\hat B$ stay 
almost disjoint. Eq.\ (\ref{so}) governs the maximum 
speed at which a local 
excitation can travel through the lattice and 
the maximum speed at which
correlations can build up over time. 

The physical setting considered here can 
equally be viewed as the study of the situation of quickly
{\it quenching} from, say, a system that is in the ground
state of some local Hamiltonian to a new local Hamiltonian
\cite{Relax,Kollath,Thermalization,Altman,Peschel,Barthel,Zurek,Simone}.
This setting has also been linked to the entanglement generation and
scaling in quenched systems \cite{Calabrese,Entanglement,Entanglement2,Dynamical,D2}.
Studies of non-equilibrium 
dynamics of quantum lattice systems of this type 
are entering a renaissance recently, not the least due to experimental 
studies becoming more and more available. With atoms
in optical lattices, for example, one can suddenly
alter the system parameters and thus observe the 
{\it dynamics of a quantum phase transition} \cite{double_well}. 
Hence, it seems only
natural to apply the machinery of Lieb-Robinson bounds to
such settings. 
However, despite
the generality of the above mentioned results on Lieb-Robinson
bounds, they are---with the exception of the recent
Refs.\ \cite{Schlein}---constrained in the sense that
they only apply to spin systems, so finite-dimensional 
constituents, quite unlike the situation encountered in
many settings of non-equilibrium dynamics. 

In this work, we derive Lieb-Robinson bounds to harmonic
lattice systems on general graphs. Such models correspond
to discrete versions of free fields, lattice vibrations,
or the superfluid limit of the Bose-Hubbard model. As harmonic
models, applicable merely to a class of systems, 
the resulting bounds can indeed be made very tight, e.g.,
for local Hamiltonians, we find 
stronger than exponential suppression, while for algebraically
decaying interactions the corrections to locality are also algebraically 
suppressed.
Within the considered class of models the very tight
connection between the approximate locality of operators 
and the locality of the Hamiltonian is thus revealed.

These systems serve as instructive theoretical laboratories for
more elaborate interacting theories for infinite-dimensional
quantum systems (and Lieb-Robinson bounds have fundamental
implications, e.g., in the context of using harmonic systems as quantum channels \cite{Dynamical,D2,Plenio}), about which 
little is known when it comes to Lieb-Robinson bounds
(see, however, Refs.\ \cite{Schlein,New} for recent progress
on bounded anharmonicities).
In this way, we continue the program  of Refs.\ \cite{Gap,Schuch}, 
building upon earlier work
primarily on clustering of correlations and {\it ``area theorems''} in
harmonic lattice systems \cite{Harmonic,Area,Area2,Botero,Frustrated,Iran}. 

This chapter is organized as follows. We first define the 
models under consideration, and explain what we mean
by having a general lattice. We then present bounds on 
the time evolved canonical coordinates and 
make the causal cone explicit. For the important class of
Weyl operators on the lattice, we explicitly find bounds on
their operator norms and discuss generalizations to
arbitrary operators. As an example, we discuss the case
of the discrete version of the Klein-Gordon field, and 
show how the approximate locality in the lattice model
becomes the exact locality in the continuum limit. 
The proofs of the main results are then presented in
a separate section in great detail. 
As such, the findings in this work complement the findings
of Ref.\ \cite{Schlein}, which considers in its harmonic
part nearest-neighbor interactions of translationally
invariant models on cubic lattices. We finally discuss
the implications to quenched many-body systems far
from equilibrium.
    
\section{Considered models and main results}

We consider harmonic systems on general lattices. Such lattices are described by
an undirected graph $G=(L,E)$, with vertices $L$ and edge set $E$. The vertices
$L$ correspond to the physical degrees of freedom, here bosonic modes with 
Hilbert space ${\cal H}_i={\cal L}^2(\rr)$, $i\in L$. Edges
reflect a neighborhood relation on the lattice.
On $L$ we use the graph-theoretical distance $dist(i,j)$ 
between $i,j\in L$, i.e., the shortest path connecting $i$ and $j$.
On such a general set of lattice sites $L$ we consider 
harmonic Hamiltonians of the form
\begin{equation}
\label{Hamiltonian}
\hat{H}=\frac{1}{2}\sum_{i,j\in L}\left(\hat{x}_iX_{i,j}\hat{x}_j+\hat{p}_iP_{i,j}\hat{p}_j\right),
\end{equation}
where $X_{i,j}=X_{j,i}\in\rr$, $P_{i,j}=P_{j,i}\in\rr$ and the $\hat{x}_i$, $\hat{p}_i$ are canonical coordinates obeying the usual commutation relations (we set $\hbar=1$) $[\hat{x}_i,\hat{x}_j]=[\hat{p}_i,\hat{p}_j]=0$, $[\hat{x}_i,\hat{p}_j]=\mi\delta_{i,j}$. 
Identifying
\begin{equation}
\label{ident}
A_{i,j}=\frac{X_{i,j}+P_{i,j}}{2},\;
B_{i,j}=\frac{X_{i,j}-P_{i,j}}{2},\;
\hat{b}_i=\frac{\hat{x}_i+\mi\hat{p}_i}{\sqrt{2}},
\end{equation}
the above Hamiltonian is equivalent to a Hamiltonian quadratic in
 the annihilation and creation operators of the
bosonic modes
($[\hat{b}_i,\hat{b}_j]=0$, $[\hat{b}_i,\hat{b}_j^\dagger]=\delta_{i,j}$)
of the form
\begin{equation}
\hat{H}=\frac{1}{2}\sum_{i,j\in L}\left(\hat{b}_i^\dagger A_{i,j}\hat{b}_j+\hat{b}_i A_{i,j}\hat{b}_j^\dagger+\hat{b}_i B_{i,j}\hat{b}_j+\hat{b}_i^\dagger B_{i,j}\hat{b}_j^\dagger\right).
\end{equation}
All the following results may thus also be obtained for the above 
Hamiltonian and bosonic operators by using the identification in Eq.\ (\ref{ident}). 
We suppose that $L$ is countable such that we may identify the 
couplings $(C_{i,j})_{i,j\in L}$, $C=X,P$, with matrices, thereby 
defining the operator norm $\|C\|$ and multiplications in the 
usual matrix sense. We denote the time evolution of
operators $\hat{o}$ in the Heisenberg picture
as
\begin{equation}
	\hat{o}(t)=\me^{\mi\hat{H}t}\hat{o}\,\me^{-\mi\hat{H}t},
\end{equation}
 and by $\|\cdot\|$ the operator norm throughout. 

\subsection{Local couplings}

In this subsection, we will derive Lieb-Robinson bounds for 
harmonic systems with arbitrary local interactions on the
graph. We will see that outside a causal cone we obtain a stronger than exponentially 
suppressed influence of time evolved canonical coordinates
in the Heisenberg picture. 
In the above notation, local means that 
\begin{equation}
X_{i,j}=P_{i,j}=0 \text{ for } dist(i,j)>R.
\end{equation}
For notational clarity we write 
\begin{equation}\label{me1}
	d_{i,j}:= dist(i,j)/R,\;\;\; \tau:=\max\{\sqrt{\|PX\|},\sqrt{\|XP\|}\}|t|
\end{equation}
in the following, 
and denote by $\lceil x\rceil=\min\{z\in\zz\, |\, x\le z\}$ the ceiling function. For a Hamiltonian as in 
Eq.\ (\ref{Hamiltonian}) with local couplings as 
above we prove the following theorems.

\begin{theorem}[Lieb-Robinson bounds for local couplings]  Writing $b_{i,j}=\lceil d_{i,j}/2\rceil$ and $a_{i,j}=\max\{0,\lceil (d_{i,j}-1)/2\rceil\}$, one has
\begin{equation}
\begin{split}
\frac{\sqrt{\|PX\|}}{\|P\|}\left\|\left[\hat{x}_i(t),\hat{x}_j\right]\right\|,
\frac{\sqrt{\|XP\|}}{\|X\|}\left\|\left[\hat{p}_i(t),\hat{p}_j\right]\right\|
&\le
\frac{\tau^{2a_{i,j}+1}\cosh\left(\tau\right)}{(2a_{i,j}+1)!},\\
\left\|\left[\hat{x}_i(t),\hat{p}_j\right]\right\|, \left\|\left[\hat{p}_i(t),\hat{x}_j\right]\right\|&\le
\frac{\tau^{2b_{i,j}}\cosh\left(\tau\right)}{(2b_{i,j})!}.
\end{split}
\end{equation}
\end{theorem}
We note that ($d\in\nn$ and we use $1/d!\le (\me/d)^dd^{-1/2} $)
\begin{equation}
\frac{\tau^d\cosh(\tau)}{d!}\le\frac{\me^{\tau+d\left(1+\log(\tau)-\log(d)\right)}}{\sqrt{d}},
\end{equation}
i.e., for sufficiently large $dist(i,j)$, one finds a faster-than-exponential
decay. In the subsequent formulation
we make this more explicit by defining a ``light cone", $\me\tau<d_{i,j}$ (i.e., $c|t|<dist(i,j)$),
with a ``speed of light" given by 
\begin{equation}\label{me2}	
	c:=\me R\max\{\|XP\|^{1/2},\|PX\|^{1/2}\}.
\end{equation}
Then commutators of  ``space-like separated" operators 
are strongly suppressed. This $c$ is an upper bound to the speed with which
a local excitation would travel 
through the lattice in a non-equilibrium
situation. This kind of argument is used, e.g., in
Ref.\ \cite{Relax}, where a central limit type argument was used
to show exact relaxation in a quenched system --
the intuition being that inside the cone excitations randomize the system while the influence of excitations outside the cone 
is negligible, which is essentially a Lieb-Robinson-type 
argument. In a very 
similar fashion, one can argue that the speed at which 
correlations build up in time is governed by these bounds
(see also Ref.\ \cite{Entanglement2}). Again, the above
speed is an upper bound to which one can correlate 
separate regions starting from an uncorrelated state under 
quenched, non-equilibrium dynamics.

As mentioned before, these bounds have immediate implication
to the evaluation of capacities of 
harmonic chains, when being used as 
convenient quantum channels. Such an idea of {\it transporting 
quantum information} through interacting quantum systems
is an appealing one, as no exact local control is required,
and the transport of quantum information is merely
due to the free evolution of an excitation under local 
dynamics.\footnote{For harmonic instances, see, e.g., Refs.\
\cite{Dynamical,Plenio,D2}, but there is a vast
literature also for spin systems and other 
finite-dimensional quantum systems, to mention all of 
which would be beyond the scope of this chapter.}
Following the argument of Ref.\ \cite{Entanglement2},
one could use such a harmonic chain as a quantum channel
through which one sends classical information, encoded
in the application of a local unitary at some site, and
letting the system freely evolve in time. The decoding 
corresponds to a readout at a distant site. Then indeed,
outside the cone defined by the ``speed of light", the
{\it classical information capacity} 
$C$ of this quantum channel
is exponentially small. This means 
that the classical information capacity of harmonic chains used as
quantum channels is -- for a fixed time -- exponentially
small in the distance between sender and observer.
The causal cone is made even more explicit in the 
subsequent formulation.

\begin{theorem}[Alternate version making the causal cone explicit] 
\label{weyl_appl} Let $\me\tau < d_{i,j}$. Then 
\begin{equation}
\begin{split}
\frac{\sqrt{\|PX\|}}{\|P\|}\left\|\left[\hat{x}_i(t),\hat{x}_j\right]\right\|,
\frac{\sqrt{\|XP\|}}{\|X\|}\left\|\left[\hat{p}_i(t),\hat{p}_j\right]\right\|,
\left\|\left[\hat{x}_i(t),\hat{p}_j\right]\right\|, \text{ and }
\left\|\left[\hat{p}_i(t),\hat{x}_j\right]\right\|
\end{split}
\end{equation}
 are all bounded from above by
\begin{equation}
\frac{\me^{d_{i,j}\log\left(\me\tau/d_{i,j}\right)}}
{\sqrt{d_{i,j}}\left(1-\left(\me\tau/d_{i,j}\right)^2\right)}.
\end{equation}
\end{theorem}

Often, $X$ and $P$ will commute, rendering the $\max$ in Eqs.\
(\ref{me1},\ref{me2})
irrelevant.  	
In many physical situations one even 
has $P_{i,j}=\delta_{i,j}$, i.e., $\tau=\sqrt{\|X\|}|t|$, in which case the 
above bounds may be improved, in particular, the 
``speed of light" improves to $c=\me R \|X\|^{1/2}/2$. For clarity,
we explicitly state the new bounds in this case, in both ways.

\begin{theorem}[Lieb-Robinson bounds for local couplings and $P=\id$] Let $a_{i,j}=\max\{0,\lceil d_{i,j}-1\rceil\}$. Then 
\begin{equation}
\begin{split}
\sqrt{\|X\|}\left\|\left[\hat{x}_i(t),\hat{x}_j\right]\right\|
&\le
\frac{\tau^{2\lceil d_{i,j}\rceil+1}\cosh\left(\tau\right)}{(2\lceil d_{i,j}\rceil+1)!}
,\\
\frac{1}{\sqrt{\|X\|}}\left\|\left[\hat{p}_i(t),\hat{p}_j\right]\right\|&\le
\frac{\tau^{2a_{i,j}+1}\cosh(\tau)}{(2a_{i,j}+1)!},\\
\left\|\left[\hat{x}_i(t),\hat{p}_j\right]\right\|, \left\|\left[\hat{p}_i(t),\hat{x}_j\right]\right\|&\le
\frac{\tau^{2\lceil d_{i,j}\rceil}\cosh\left(\tau\right)}{(2\lceil d_{i,j}\rceil)!}.
\end{split}
\end{equation}
\end{theorem}
\begin{theorem}[Alternate version making the causal cone explicit for $P=\id$] \label{field_theorem} For $\me\tau< 2d_{i,j}$ one has
\begin{equation}
\sqrt{\|X\|}\left\|\left[\hat{x}_i(t),\hat{x}_j\right]\right\|, \left\|\left[\hat{x}_i(t),\hat{p}_j\right]\right\|, 
\left\|\left[\hat{p}_i(t),\hat{x}_j\right]\right\|
\le
\frac{\me^{2d_{i,j}\log\left(\me\tau/(2d_{i,j})\right)}}{\sqrt{d_{i,j}}\left(1-\left(\me\tau/(2d_{i,j})\right)^2\right)},
\end{equation}
and for $\me\tau < (2a_{i,j}+1)$ 
\begin{equation}
\frac{1}{\sqrt{\|X\|}}\left\|\left[\hat{p}_i(t),\hat{p}_j\right]\right\|\le
\frac{\me^{2a_{i,j}\log\left(\me\tau/(2a_{i,j}+1)\right)}}{\sqrt{a_{i,j}}\left(1-\left(\me\tau/(2a_{i,j}+1)\right)^2\right)},
\end{equation}
where now $a_{i,j}=\max\{0,\lceil d_{i,j}-1\rceil\}$.
\end{theorem}

\subsection{Application: Non-relativistic quantum mechanics yields causality in the field limit}

This section forms an application of the previous considerations. We will see
how the exact light cone of the free field is recovered from the approximate light
cone in the Lieb-Robinson theorem in the continuum limit of the lattice version of
the field theory. It is very instructive indeed to see how the tails in the superexponentially
suppressed region outside the light cone becomes more and more suppressed in this
limit. The role of the Lieb-Robinson velocity is hence taken over
by the speed of light in the relativistic model. 

We start from the {\it Klein-Gordon Hamiltonian} on $V=[0,1]^{\times D}$ in 
units $\hbar=c=1$,
\begin{equation}
\hat{H}=\frac{1}{2}\int_V\md\vec{x}\,\left[\hat{\pi}^2(\vec{x})+\sum_{d=1}^D\left(\partial_{x_d}\hat{\varphi}(\vec{x})\right)^2+m^2\hat{\varphi}^2(\vec{x})\right],
\end{equation}
where the field operators fulfill the usual commutation relations
\begin{equation}
\left[\hat{\varphi}(\vec{x}),\hat{\pi}(\vec{y})\right]=\mi\delta(\vec{x}-\vec{y}),\;\;\;
\left[\hat{\varphi}(\vec{x}),\hat{\varphi}(\vec{y})\right]=\left[\hat{\pi}(\vec{x}),\hat{\pi}(\vec{y})\right]=0.
\end{equation}
Discretizing according to $\vec{x}=\vec{i}/N$, $\vec{i}\in \{\vec{j}\in \nn^D\,|\, i_d=1,2,...,N\}=:L$, \begin{equation}
\int_V\md\vec{x}\,\hat{f}(\vec{x})\rightarrow \frac{1}{N^D} \sum_{\vec{i}\in L} \hat{f}(\vec{i}/N),\;\;
(\partial_{x_d}\hat{f})(\vec{x})\rightarrow\frac{\hat{f}(\vec{x}+\vec{n}_d/N)-\hat{f}(\vec{x})}{1/N},
\end{equation}
where $\vec{n}_d$ denotes a unit vector in direction $d$, we find 
(equipping $L$ with periodic boundary conditions),
\begin{equation}
\begin{split}
\hat{H}&\rightarrow\frac{1}{2N^D}\sum_{\vec{i}\in L}\left[\hat{\pi}^2(\vec{i}/N)+\sum_{d=1}^D
\left(\frac{\hat{\varphi}(\vec{i}/N+\vec{n}_d/N)-\hat{\varphi}(\vec{i}/N)}{1/N}\right)^2
+m^2\hat{\varphi}^2(\vec{i}/N)\right]\\
&=\frac{1}{2N^D}\left(\sum_{\vec{i}\in L}\Bigl[\hat{\pi}^2(\vec{i}/N)+\left(m^2+2DN^2\right)\hat{\varphi}^2(\vec{i}/N)\Bigr]
-N^2\!\!\!\!\!\!\sum_{\substack{\vec{i},\vec{j}\in L\\ dist(\vec{i},\vec{j})=1}}\!\!\!\!\!\!\hat{\varphi}(\vec{i}/N)\hat{\varphi}(\vec{j}/N)\right)\\
&=:\hat{H}_N.
\end{split}
\end{equation}
Then $N\rightarrow\infty$ is the valid continuum limit for a fixed $V=[0,1]^{\times D}$. Now,
\begin{equation}
\hat{x}_{\vec{i}}:=N^{-D/2}\hat{\varphi}(\vec{i}/N),\;\;\;
\hat{p}_{\vec{i}}:=N^{-D/2}\hat{\pi}(\vec{i}/N),
\end{equation}
define harmonic position and momentum operators
satisfying the canonical commutation relations, 
in terms of which we find
\begin{equation}
\hat{H}_N=\frac{1}{2}\sum_{\vec{i},\vec{j}\in L}\Bigl[\hat{p}_{\vec{i}}P_{\vec{i},\vec{j}}\hat{p}_{\vec{j}}+
\hat{x}_{\vec{i}}X_{\vec{i},\vec{j}}\hat{x}_{\vec{j}}\Bigr],
\end{equation}
where $P_{\vec{i},\vec{j}}=\delta_{\vec{i},\vec{j}}$ and
\begin{equation}
X_{\vec{i},\vec{j}}=\left(m^2+2DN^2\right)\delta_{\vec{i},\vec{j}}
-N^2\delta_{dist(\vec{i},\vec{j}),1}.
\end{equation}
We are interested in the discretized version of the commutator $[\hat{\varphi}(\vec{x},t),\hat{\varphi}(\vec{0},0)]$, which is given by
\begin{equation}
N^D
\left[\hat{x}_{\vec{i}}(t),\hat{x}_{\vec{0}}\right],
\end{equation}
and set out to apply Theorem \ref{field_theorem}. We have $R=1$ and assume w.l.o.g. that $0\le i_d\le N/2$, i.e., $d_{\vec{i},\vec{0}}=dist(\vec{i},\vec{0})=\sum_{d=1}^Di_d=N\sum_{d=1}^Dx_d\ge N|\vec{x}|$, with $|\vec{x}|$ being the euclidean norm. Furthermore, as we assumed translational invariance the eigenvalues $\lambda_{\vec{k}}$ of $X$ are given by
\begin{equation}
\lambda_{\vec{k}}=m^2+2DN^2-2N^2\sum_{d=1}^D\cos\left(2\pi k_d/N\right), 
\end{equation}
i.e., $\|X\|\le m^2+4DN^2$ (for even $N$ we have equality). 

Now fix $|t|$ and $|\vec{x}|$ such that
\begin{equation}
\me\sqrt{D}|t|<|\vec{x}|.
\end{equation}
 We then 
take the limit $N\rightarrow\infty$ such that $\sum_{d=1}^Dx_d=d_{\vec{i},\vec{0}}/N=const.$ is fulfilled for all $N$ (e.g., $\vec{x}=(1/4,0,\dots,0)$ fixes $\vec{i}=(N/4,0,\dots,0)$ and $N/4\in \nn$). Now let
$N_0\in \nn$ be such that
\begin{equation}
1>\frac{\me |t|\sqrt{D}}{|\vec{x}|}\sqrt{\frac{m^2}{4DN_0^2}+1}=:z,
\end{equation}
which yields
\begin{equation}
\me\tau<
|\vec{x}|\frac{\sqrt{\|X\|}}{\sqrt{\frac{m^2}{4N_0^2}+D}}
\le 2d_{\vec{i},\vec{0}}\frac{\sqrt{\|X\|}}{N\sqrt{\frac{m^2}{N_0^2}+4D}}<2d_{\vec{i},\vec{0}}
\end{equation}
for all $N>N_0$. This enables us to apply Theorem \ref{field_theorem} to find
\begin{equation}
\left\|\left[\hat{x}_{\vec{i}}(t),\hat{x}_{\vec{0}}\right]\right\|\le \frac{\me^{2N|\vec{x}|\log\left(z\right)}}{\sqrt{N\|X\||\vec{x}|}\left(1-z^2\right)}
\end{equation}
for all $N>N_0$,
i.e.,
\begin{equation}\label{lc}
\lim_{N\rightarrow\infty}N^D
\left\|\left[\hat{x}_{\vec{i}}(t),\hat{x}_{\vec{0}}\right]\right\|=0
\end{equation}
independent of $m$. Eq.\ (\ref{lc}) shows
 that the approximate
light cone of the Lieb-Robinson bound becomes an exact
light cone in the continuum limit. The exponentially
suppressed tails vanish, and approximate locality is
replaced by an exact locality. It is interesting to see
how this concept emerges from the bounds to 
the speed of information propagation in the sense
of Lieb-Robinson.

\begin{figure}[h!]
\begin{center}
\includegraphics[width=\textwidth]{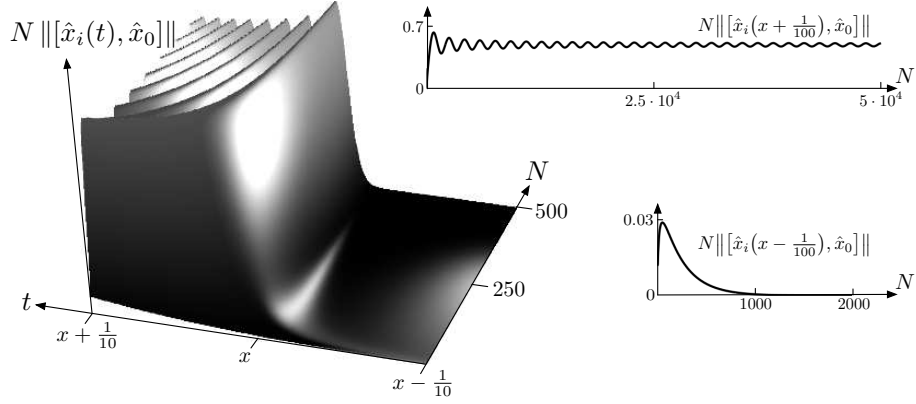}
\caption{\label{figure} The light cone in the field limit of the discrete Klein-Gordon field: Depicted is $N\|[\hat{x}_i(t),\hat{x}_0]\|$ as a funtion of $t$ and $N$. This is the discrete version of $\|[\hat{\varphi}(x,t),\hat{\varphi}(0,0)]\|$, $x=i/N$, in one spatial dimension. For $|t|<|x|$ we find for finite system size $N$ the Lieb-Robinson exponential decay in $|x|$ and for $N\rightarrow\infty$ the commutator $N\|[\hat{x}_i(t),\hat{x}_0]\|$ goes to zero for space-like separations, thus recovering exact causality.}
\end{center}
\end{figure}

The bound in Theorem \ref{field_theorem} is not quite
strong enough to recover the exact prefactor of the 
light cone $|t|<|\vec{x}|$. 
This is mainly due to the fact that we allowed 
for general lattices in the Lieb-Robinson bound. 
Demanding translational  invariance would allow for slightly 
stronger bounds. 
In Fig.\ \ref{figure} we depict exact numerical results for this 
geometrical setting in $D=1$. 

\subsection{Non-local couplings}

The previous section allowed for arbitrary local interactions.
In this section we will turn to strongly decaying 
{\it non-local couplings} of the form
\begin{equation}
\left|X_{i,j}\right|, \left|P_{i,j}\right|\le\frac{c_0}{\left(dist(i,j)+1\right)^\eta}.
\end{equation}
We define the spatial dimension of $L$ in the usual 
sense: For all spheres $S_r(i)\subset L$ with radius $r\in\nn$ centered at $i\in L$, 
\begin{equation}
S_r(i)=\left\{k\in L\big|dist(k,i)=r\right\},
\end{equation} 
there exists a smallest $D>0$ such that for all $0<r\in\nn$
\begin{equation}
\sup_{i\in L}\left|S_r(i)\right|\le c_Dr^{D-1}
\end{equation}
for some constant $c_D>0$. This number $D$ is taken as the dimension of the lattice. 
We find that the decay of interactions is inherited by the decay of the
operator norm of the commutator of canonical coordinates. The same power in the exponent
as in the interaction again appears in the Lieb-Robinson bounds. Note the
(not accidental) similarity with the inheritance of the decay of correlation
functions dependent on the decay of interactions in Ref.\ \cite{Gap}.

\begin{theorem}[Bounds for non-local couplings] 
\label{t5}
Let $\eta>D$. Then
\begin{equation}
\begin{split}
\left\|\left[\hat{x}_i(t),\hat{x}_j\right]\right\|,\left\|\left[\hat{p}_i(t),\hat{p}_j\right]\right\|&\le\frac{\sinh(\tau)}{a_0(1+dist(i,j))^{\eta}},\\
\left\|\left[\hat{x}_i(t),\hat{p}_j\right]\right\|, \left\|\left[\hat{p}_i(t),\hat{x}_j\right]\right\|&\le
\delta_{i,j}+\frac{\cosh(\tau)}{a_0(1+dist(i,j))^{\eta}},
\end{split}
\end{equation}
where $\tau=a_0c_0|t|$, $a_0=c_D2^{\eta+1}\zeta(1-D+\eta)$, and $\zeta$ is the Riemann zeta function.
\end{theorem}

\subsection{Weyl operators}

A class of operators that play a central role in harmonic
systems are the Weyl operators. Denoting the support of a Weyl operator $\hat{W}_{\xi}$ as
$\Xi\subset L$ it may be written as
\begin{equation}
\hat{W}_{\xi}=\me^{\mi\sum_{i\in \Xi}\left(p_i\hat{x}_i-x_i\hat{p}_i\right)},
\text{ where } \xi=(x_1,...,x_{|\Xi|},p_1,...,p_{|\Xi|})\in \rr^{2|\Xi|}.
\end{equation}
Via the Fourier-Weyl relation general bounded operators may be expressed in terms of these operators, see below.

We define the distance of two sets $A,B\subset L$ as
\begin{equation}
dist(A,B)=\min_{i\in A, j\in B}dist(i,j),
\end{equation}
and the surface area of a set $A\subset L$ as $|\partial A|$, where
\begin{equation}
\partial A=\left\{i\in A\,\big|\,\exists \,j\in L\backslash A: dist(i,j)=1\right\}
\end{equation}
defines the set of lattice sites on the surface of $A$.
The following theorem establishes a connection between 
commutators of Weyl operators and previously derived bounds 
on the canonical coordinates. Note in the subsequent theorem
the dependence on the right hand side of the
operator norms $||\xi||,||\xi'||$ of $\xi$ and $\xi'$, 
whereas on the left hand side of Eq.\ (\ref{big}), 
we have the operator norm for commutators of 
Weyl operators.

\begin{theorem}[Lieb-Robinson bounds for Weyl operators] Let 
\begin{equation}
\hat{W}_{\xi}=\me^{\mi\sum_{i\in \Xi}\left(p_i\hat{x}_i-x_i\hat{p}_i\right)},\;\;\;
\hat{W}_{\xi^\prime}=\me^{\mi\sum_{i\in \Xi^\prime}\left(p_i\hat{x}_i-x_i\hat{p}_i\right)}
\end{equation}
be Weyl operators as defined above.
Then
\begin{equation}\label{big}
\begin{split}
\left\|
\left[\hat{W}_{\xi}(t),\hat{W}_{\xi^\prime}\right]\right\|
\le &
\|\xi\|\|\xi^\prime\|
\sum_{i\in \Xi,j\in \Xi^\prime}\big(
\left\|\left[x_i(t),x_{j}\right]\right\|
+\left\|\left[x_i(t),p_{j}\right]\right\| \\
& \hspace{3.5cm}+\left\|\left[p_i(t),x_{j}\right]\right\|+
\left\|\left[p_i(t),p_{j}\right]\right\|
\big)\\
\le & \,c_D \|\xi\|\|\xi^\prime\|\min\left\{\left|\partial\Xi\right|,\left|\partial\Xi^\prime\right|\right\}\\
&\times
\sum_{d=dist(\Xi,\Xi^\prime)}^\infty f(d)d^{D-1}\left(1+c_D(d-dist(\Xi,\Xi^\prime))^D\right).
\end{split}
\end{equation}
where $f:\nn\rightarrow\rr$ is a function such that
\begin{equation}\label{dec}
\left\|\left[x_i(t),x_{j}\right]\right\|
+\left\|\left[x_i(t),p_{j}\right]\right\|+\left\|\left[p_i(t),x_{j}\right]\right\|+
\left\|\left[p_i(t),p_{j}\right]\right\|\le f(dist(i,j)).
\end{equation}
Employing, e.g., Theorem \ref{weyl_appl}, we have for 
$\me\tau <dist(\Xi,\Xi^\prime)/R=:d_{\Xi,\Xi^\prime}$ that
\begin{equation}
\begin{split}
\left\|\left[\hat{W}_{\xi}(t),\hat{W}_{\xi^\prime}\right]\right\|
\le C \min\left\{\left|\partial\Xi\right|,\left|\partial\Xi^\prime\right|\right\}g\left(\frac{\me\tau }{d_{\Xi,\Xi^\prime}}\right)
\me^{d_{\Xi,\Xi^\prime}\log\left(\me\tau /d_{\Xi,\Xi^\prime}\right)}
d^{D-3/2}_{\Xi,\Xi^\prime},
\end{split}
\end{equation}
where 
\begin{equation}
C=R^{D-1}c_D \|\xi\|\|\xi^\prime\|\left(\frac{\|P\|}{\sqrt{\|PX\|}}+\frac{\|X\|}{\sqrt{\|XP\|}}+2\right)
\end{equation}
and
the function $g:(0,1)\rightarrow \rr$,
\begin{equation}
g(z)=\frac{1}{1-z^2}\sum_{d=0}^\infty z^{d/R}(d+1)^{D-1}(1+c_D(d+1)^{D})\ge 0,
\end{equation}
is increasing in $z$ with $\lim_{z\rightarrow 0}=1$.
\end{theorem}

Note that we have in 
Eq.\ (\ref{dec})
expressed this statement in terms of a function $f$
that grasps the decay of operator norms of commutators of canonical 
coordinates. Whenever one can identify such a function, e.g., through 
theorems 1-5, a result on Weyl
operators can be deduced. Needless to say, in the same way we have
applied Theorem \ref{weyl_appl}, we could have
made use of Theorem \ref{t5}: Essentially, the decay in operator norms
of canonical coordinates is inherited by the expression for Weyl operators.
Due to the sum in Eq.\ (\ref{big}), however, it can 
happen that no decay follows for Weyl operators if the (i) dimension
of the lattice is too high or (ii) the decay of $f$ is too slow. For finite-dimensional
lattices, however, sufficiently fast algebraically decaying (i.e., sufficiently large $\eta$)
interactions yield an algebraic Lieb-Robinson-type
statement for the commutator of two Weyl operators.

Note also that only the surface areas of the two sets $\Xi,\Xi'$ enter the 
bound, but not the cardinality of the support. This allows for 
infinite regions (for $D=1$ both may be supported on infinite intervals,
for $D>1$ only one of $\Xi,\Xi'$ needs to have a finite surface area) separated by $dist(\Xi,\Xi')$.
\subsection{More general operators}
A general bounded operator $\hat{o}$ supported on $\Xi\subset L$ may be expressed as
\begin{equation}
\hat{o}=\frac{1}{(2\pi)^{|\Xi |}}\int_{\rr^{2|\Xi |}}\!\!\!\!\!\!\!\!\!\!\md\xi\;\chi_{\hat{o}}(-\xi)\hat{W}_{\xi},
\end{equation}
where
\begin{equation}
\chi_{\hat{o}}(\xi)=\text{tr}\left[\hat{o}\hat{W}_{\xi}\right]
\end{equation}
is the characteristic function of $\hat{o}$. This allows to deduce bounds for general bounded operators using the bounds on Weyl operators stated above:
\begin{equation}
\left\|
\left[\hat{o}(t),\hat{o}^\prime\right]\right|\le
\frac{1}{(2\pi)^{|\Xi |+|\Xi^\prime |}}
\int_{\rr^{2|\Xi |}}\!\!\!\!\!\!\!\!\!\!\md\xi\;\int_{\rr^{2|\Xi^\prime |}}\!\!\!\!\!\!\!\!\!\!\md\xi^\prime\;\left|\chi_{\hat{o}}(-\xi)\chi_{\hat{o}^\prime}(-\xi^\prime)\right|
\left\|
\left[\hat{W}_{\xi}(t),\hat{W}_{\xi^\prime}\right]\right\|.
\end{equation}

Bounds for more general, possibly unbounded, operators that are finite sums of finite
products of canonical operators (or bosonic creation and annihilation operators) may be obtained by repeatedly 
employing operator identities such as
\begin{equation}
\left[\hat{A}\hat{B},\hat{C}\right]=\hat{A}\left[\hat{B},\hat{C}\right]+\left[\hat{A},\hat{C}\right]\hat{B},\;\;
\left[\hat{A},\hat{B}\hat{C}\right]=\left[\hat{A},\hat{B}\right]\hat{C}+\hat{B}\left[\hat{A},\hat{C}\right],
\end{equation}
e.g., bosonic density-density commutators may be written as
\begin{equation}
\begin{split}
\left[\hat{n}_i(t),\hat{n}_j\right]&=\hat{b}_i^\dagger(t)\left[\hat{b}_i(t),\hat{n}_j\right]+\left[\hat{b}_i^\dagger(t),\hat{n}_j\right]\hat{b}_i(t)\\
&=
\hat{b}_i^\dagger(t)\left(
\left[\hat{b}_i(t),\hat{b}_j^\dagger\right]
\hat{b}_j
+\hat{b}_j^\dagger\left[\hat{b}_i(t),\hat{b}_j\right]
\right)\\
&\hspace{2cm}+
\left(
\left[\hat{b}_i^\dagger(t),\hat{b}^\dagger_j\right]\hat{b}_j
+\hat{b}^\dagger_j\left[\hat{b}_i^\dagger(t),\hat{b}_j\right]
\right)\hat{b}_i(t),
\end{split}
\end{equation}
which yields
\begin{equation}
\begin{split}
\left|\left\langle\left[\hat{n}_i(t),\hat{n}_j\right]
\right\rangle\right|&\le
\left|\left\langle
\hat{b}_i^\dagger(t)
\hat{b}_j
\right\rangle\right|\left\|\left[\hat{b}_i(t),\hat{b}_j^\dagger\right]\right\|
+
\left|\left\langle\hat{b}_i^\dagger(t)\hat{b}_j^\dagger\right\rangle\right|\left\|\left[\hat{b}_i(t),\hat{b}_j\right]\right\|\\
&\hspace{1cm}+
\left|\left\langle\hat{b}_j\hat{b}_i(t)\right\rangle\right|\left\|\left[\hat{b}_i^\dagger(t),\hat{b}^\dagger_j\right]\right\|
+\left|\left\langle\hat{b}^\dagger_j\hat{b}_i(t)\right\rangle\right|\left\|\left[\hat{b}_i^\dagger(t),\hat{b}_j\right]\right\|,
\end{split}
\end{equation}
where bounds on the commutators may then be obtained by identifying bosonic operators by canonical
operators through Eq.\ (\ref{ident}) and employing the above derived bounds.


\section{Proofs}
In this section, we will present in detail
the proofs of the previous statements.
\subsection{Preliminaries}

We write the Hamiltonian in Eq.\ (\ref{Hamiltonian}) as
\begin{equation}
	\hat{H}=\frac{1}{2}\sum_{n_i,n_j=1}^{2|L|}\hat{r}_{n_i}H_{n_i,n_j}\hat{r}_{n_j},
\end{equation}
where we have arranged lattice sites such that $H_{n_i,n_j}=X_{i,j}$ ($=P_{i,j}$) for $1\le n_i,n_j \le |L|$
($L+1\le n_i,n_j \le 2|L|$) and $\hat{r}_{n_i}=\hat{x}_{i}$ ($=\hat{p}_{i}$) for $1\le n_i \le |L|$
($L+1\le n_i \le 2|L|$). 
Now consider the time evolution of the operator
\begin{equation}
\hat{r}_{n_i}(t):=\me^{\mi \hat{H}t}\hat{r}_{n_i}\me^{-\mi \hat{H}t}.
\end{equation}
By solving Heisenberg's equation of motion or, alternatively, by employing the Baker-Hausdorff formula,
one finds
\begin{equation}
\hat{r}_{n_i}(t)=\sum_{n_j=1}^{2|L|}\left(\me^{-\sigma Ht}\right)_{n_i,n_j}\hat{r}_{n_j},\text{ where }
\sigma_{n_i,n_j}=\mi\left[\hat{r}_{n_i},\hat{r}_{n_j}\right].
\end{equation}
This yields for the commutator
\begin{equation}
\mi\left[\hat{r}_{n_i}(t),\hat{r}_{n_j}\right]=
\mi\sum_{n_k=1}^{2|L|}\left(\me^{-\sigma Ht}\right)_{n_i,n_k}\left[\hat{r}_{n_k},\hat{r}_{n_j}\right]
=\left(\me^{-\sigma Ht}\sigma\right)_{n_i,n_j}.
\end{equation}
Now, separating the terms with an even power in $n$ 
from the terms with an odd power, we get
\begin{equation}
\begin{split}
\me^{-\sigma Ht}&=\sum_{n=0}^\infty\frac{t^n}{n!}\left(\begin{array}{cc}0&P\\ -X&0\end{array}\right)^n\\
&=\sum_{n=0}^\infty\frac{t^{2n+1}}{(2n+1)!}\left(\begin{array}{cc}0&P\\ -X&0\end{array}\right)^{2n+1}
+\sum_{n=0}^\infty\frac{t^{2n}}{(2n)!}\left(\begin{array}{cc}0&P\\ -X&0\end{array}\right)^{2n}\\
&=\sum_{n=0}^\infty\frac{(-1)^nt^{2n+1}}{(2n+1)!}\left(\begin{array}{cc}\left(PX\right)^n&0\\ 0&\left(XP\right)^n\end{array}\right)
\left(\begin{array}{cc}0&P\\ -X&0\end{array}\right)\\
&\hspace{3cm}+\sum_{n=0}^\infty\frac{(-1)^nt^{2n}}{(2n)!}\left(\begin{array}{cc}\left(PX\right)^n&0\\ 0&\left(XP\right)^n\end{array}\right).
\end{split}
\end{equation}
Hence, 
\begin{equation}
\label{the_commutators}
\begin{split}
\mi\left[\hat{x}_i(t),\hat{x}_j\right]&= 
\sum_{n=0}^\infty\frac{(-1)^nt^{2n+1}}{(2n+1)!}\left((PX)^nP\right)_{i,j}=: \id\cdot C^{xx}_{i,j}(t),\\
\mi\left[\hat{p}_i(t),\hat{p}_j\right]&= 
\sum_{n=0}^\infty\frac{(-1)^nt^{2n+1}}{(2n+1)!}\left((XP)^nX\right)_{i,j}=: \id\cdot  C^{pp}_{i,j}(t),\\
\mi\left[\hat{x}_i(t),\hat{p}_j\right]&=- 
\sum_{n=0}^\infty\frac{(-1)^nt^{2n}}{(2n)!}\left((PX)^n\right)_{i,j}=:  \id\cdot C^{xp}_{i,j}(t),\\
\mi\left[\hat{p}_i(t),\hat{x}_j\right]&=
\sum_{n=0}^\infty\frac{(-1)^nt^{2n}}{(2n)!}\left((XP)^n\right)_{i,j}=:  \id\cdot C^{px}_{i,j}(t).
\end{split}
\end{equation}
These expressions will form the starting point of the subsequent considerations.

\subsection{Local couplings}

We will need the following lemma. It states that finite 
powers of local coupling matrices defined on graphs 
remain local couplings, albeit with a larger range.

\begin{lemma}[Products of local couplings]
Let $A=(A_{i,j})_{i,j\in L}$ be such that $A_{i,j}=0$ for $dist(i,j)>R$. Then for $n\in\nn$
\begin{equation}
(A^n)_{i,j}=0 \text{ for all } i,j\in L \text{ with } dist(i,j)>nR.
\end{equation}
\end{lemma}
\proof For $n=1$ the statement is obviously true. Now let $(A^n)_{i,j}=0$ for $dist(i,j)>nR$. Then
\begin{equation}
\left(A^{n+1}\right)_{i,j}=\sum_{k\in L}\left(A^n\right)_{i,k}A_{k,j}.
\end{equation}
Now let $k\in L$. If $dist(k,j)>R$ then this $k$ does not contribute to the sum
as $A_{k,j}=0$. Now let $dist(i,j)>(n+1)R$. Then we have that also if $dist(k,j)\le R$ it does not contribute to the sum
as then $dist(i,k)>nR$ and therefore $(A^n)_{i,j}=0$:  
\begin{equation}
	(n+1)R<dist(i,j)\le dist(i,k)+dist(k,j)\le dist(i,k)+R,
\end{equation}
i.e, $dist(i,k)>nR$. \qed

Thus, if we have $X_{i,j}=P_{i,j}=0$ for $dist(i,j)>R$, we may write (see Eqs.\ (\ref{the_commutators})),
\begin{equation}
\left|C_{i,j}^{xx}(t)\right|\le
\sum_{n=a_{i,j}}^\infty\frac{|t|^{2n+1}}{(2n+1)!}\left|\left((PX)^nP\right)_{i,j}\right|,
\end{equation}
where $a_{i,j}=\max\{0,\lceil (d_{i,j}-1)/2\rceil\}$ and we recall that $d_{i,j}=dist(i,j)/R$. As one has for any matrix that $|M_{i,j}|\le \|M\|$, we find
\begin{equation}
\left|C_{i,j}^{xx}(t)\right|\le\frac{\|P\|}{\sqrt{\|PX\|}}
\sum_{n=a_{i,j}+1/2}^\infty\frac{\tau^{2n}}{(2n)!},
\end{equation}
where we recall that $\tau=\max\{\sqrt{\|PX\|},\sqrt{\|XP\|}\}|t|$. Similarly ($b_{i,j}:=\lceil d_{i,j}/2\rceil$)
\begin{equation}
\begin{split}
\left|C_{i,j}^{pp}(t)\right|&\le
\frac{\|X\|}{\sqrt{\|XP\|}}
\sum_{n=a_{i,j}+1/2}^\infty\frac{\tau^{2n}}{(2n)!}\\
\left|C_{i,j}^{xp}(t)\right|, \left|C_{i,j}^{px}(t)\right|&\le
\sum_{n=b_{i,j}}^\infty\frac{\tau^{2n}}{(2n)!}.
\end{split}
\end{equation}
In the case of $P_{i,j}=\delta_{i,j}$ these bounds read
\begin{equation}
\begin{split}
\left|C_{i,j}^{xx}(t)\right|&\le\frac{1}{\sqrt{\|X\|}}
\sum_{n=\lceil d_{i,j}\rceil+1/2}^\infty\frac{\tau^{2n}}{(2n)!}\\
\left|C_{i,j}^{pp}(t)\right|&\le
\sqrt{\|X\|}
\sum_{n=\max\{0,\lceil d_{i,j}-1\rceil\}+1/2}^\infty\frac{\tau^{2n}}{(2n)!}\\
\left|C_{i,j}^{xp}(t)\right|, \left|C_{i,j}^{px}(t)\right|&\le
\sum_{n=\lceil d_{i,j}\rceil}^\infty\frac{\tau^{2n}}{(2n)!}.
\end{split}
\end{equation}
Now, for $c\ge 0$,
\begin{equation}
\sum_{n=c}^\infty\frac{\tau^{2n}}{(2n)!}=\tau^{2c}\sum_{n=0}^\infty\frac{\tau^{2n}}{(2n+2c)!}
\le \frac{\tau^{2c}}{(2c)!}\sum_{n=0}^\infty\frac{\tau^{2n}}{(2n)!}=\frac{\tau^{2c}}{(2c)!}\cosh(\tau),
\end{equation}
also, if $\me\tau < 2c $, we find
\begin{equation}
\sum_{n=c}^\infty\frac{\tau^{2n}}{(2n)!}\le\sum_{n=c}^\infty\left(\frac{\me\tau}{2n}\right)^{2n}(2n)^{-1/2}
\le \frac{1}{\sqrt{2c}}\sum_{n=c}^\infty\left(\frac{\me\tau}{2c}\right)^{2n}
= \frac{\left(\me\tau / 2c\right)^{2c}}{\sqrt{2c}\left(1-(\me\tau / 2c)^2\right)},
\end{equation}
where we have used that $n!\ge (n/\me)^nn^{1/2}$ for $n\ge 1$.

\subsection{Non-local couplings}
Let $|M_{i,j}|\le [1+dist(i,j)]^{-\eta}$. For such couplings we have ($d_{i,j}:=dist(i,j)$)
\begin{equation}
\begin{split}
\left|\left(M^2\right)_{i,j}\right|&\le(1+d_{i,j})^{-\eta}\sum_{k}
\left(\frac{1+d_{i,j}}{(1+d_{i,k})(1+d_{k,j})}\right)^\eta
\\
&\le
(1+d_{i,j})^{-\eta}\sum_{k}
\left(\frac{1+d_{i,k}+1+d_{k,j}}{(1+d_{i,k})(1+d_{k,j})}\right)^\eta
\\
&\le
\left(\frac{2}{1+d_{i,j}}\right)^{\eta}\sum_{k}
\left(\frac{1+\max\{d_{i,k},d_{k,j}\}}{(1+d_{i,k})(1+d_{k,j})}\right)^\eta,
\end{split}
\end{equation}
where we have used the triangle inequality and $(a+b)\le 2\max\{a,b\}$.
Now, 
\begin{equation}
(1+\max\{d_{i,k},d_{k,j}\})^\eta\le(1+d_{i,k})^\eta+(1+d_{k,j})^\eta,
\end{equation}
i.e.,
an upper bound for the above sum over $k$ is 
given by
\begin{equation}
\begin{split}
\sum_{k}
\left(
\frac{1}{(1+d_{i,k})^\eta}+\frac{1}{(1+d_{k,j})^\eta}\right)
&\le 2\sup_{i\in L}\sum_{k}
\frac{1}{(1+d_{i,k})^\eta}\\
&=
2
\sum_{r=0}^\infty \frac{\sup_{i\in L}\sum_{k}\delta_{r,d_{i,k}}}{(1+r)^\eta}
,
\end{split}
\end{equation}
where
\begin{equation}
\sum_{k}\delta_{r,d_{i,k}}=\left|\left\{k\in L\big|d_{k,i}=r\right\}\right|
=|S_r(i)|,
\end{equation}
which we may bound using the definition of the dimension of the graph to find
\begin{equation}
\left|\left(M^2\right)_{i,j}\right|\le c_D\frac{2^{\eta+1}}{(1+d_{i,j})^{\eta}}
\sum_{r=0}^\infty \frac{1}{(1+r)^{\eta-D+1}},
\end{equation}
which converges if $\eta>D$, in which case we have
\begin{equation}
\left|\left(M^2\right)_{i,j}\right|\le \frac{a_0}{(1+d_{i,j})^{\eta}},\;\;\;
a_0=c_D2^{\eta+1}\zeta(1-D+\eta),
\end{equation}
where $\zeta$ is the Riemann zeta function. By induction we then find
\begin{equation}
\left|\left(X^n\right)_{i,j}\right|,\left|\left(P^n\right)_{i,j}\right|\le \frac{c_0^{n}a_0^{n-1}}{(1+d_{i,j})^{\eta}}
\end{equation}
for $n\ge 1$, implying (recalling that $\tau=c_0a_0|t|$)
\begin{equation}
\begin{split}
\left|C_{i,j}^{xx}(t)\right|&\le
\frac{1}{a_0(1+dist(i,j))^{\eta}}
\sum_{n=0}^\infty\frac{\tau^{2n+1}}{(2n+1)!}=\frac{\sinh(\tau)}{a_0(1+dist(i,j))^{\eta}},\\
\left|C_{i,j}^{xp}(t)\right|&\le
\delta_{i,j}+
\frac{1}{a_0(1+dist(i,j))^{\eta}}
\sum_{n=0}^\infty\frac{\tau^{2n}}{(2n)!}=\delta_{i,j}+\frac{\cosh(\tau)}{a_0(1+dist(i,j))^{\eta}},
\end{split}
\end{equation}
and similarly
\begin{equation}
\begin{split}
\left|C_{i,j}^{pp}(t)\right|&\le
\frac{\sinh(\tau)}{a_0(1+dist(i,j))^{\eta}},\\
\left|C_{i,j}^{px}(t)\right|&\le
\delta_{i,j}+\frac{\cosh(\tau)}{a_0(1+dist(i,j))^{\eta}}.
\end{split}
\end{equation}
\subsection{Weyl operators}
For operators $\hat{W}_{\xi}$ as above we find
\begin{equation}
\hat{W}_{\xi}(t)=\me^{\mi\sum_{i\in \Xi}\left(p_i\hat{x}_i(t)-x_i\hat{p}_i(t)\right)},\;\;\;\xi=(x_1,...,x_{|\Xi|},p_1,...,p_{|\Xi|})\in \rr^{2|\Xi|}
\end{equation}
Employing the Baker-Hausdorff identity we then have, see Eq.\ (\ref{the_commutators}),
\begin{equation}
\hat{W}_{\xi}(t)\hat{W}_{\xi^\prime}=\hat{W}_{\xi^\prime}\hat{W}_{\xi}(t)\me^{\mi\sum_{i\in\Xi,j\in \Xi^\prime}\left(
p_ip_j^\prime C^{xx}_{i,j}(t)-p_ix_j^\prime C^{xp}_{i,j}(t)
-x_ip_j^\prime C^{px}_{i,j}(t)+
x_ix_j^\prime C^{pp}_{i,j}(t)
\right)},
\end{equation}
i.e.,
\begin{equation}
\begin{split}
\left\|
\left[\hat{W}_{\xi}(t),\hat{W}_{\xi^\prime}\right]\right\|
&\le\left\|\me^{\mi\sum_{i\in\Xi,j\in \Xi^\prime}\left(
p_ip_j^\prime C^{xx}_{i,j}(t)-p_ix_j^\prime C^{xp}_{i,j}(t)
-x_ip_j^\prime C^{px}_{i,j}(t)+
x_ix_j^\prime C^{pp}_{i,j}(t)
\right)}-\id\right\|\\
&\le\|\xi\|\|\xi^\prime\|
\sum_{i\in \Xi,j\in \Xi^\prime}\left(
\left|C^{xx}_{i,j}(t)\right|
+\left|C^{xp}_{i,j}(t)\right|
+\left|C^{px}_{i,j}(t)\right|+
\left|C^{pp}_{i,j}(t)\right|
\right)\\
&=\|\xi\|\|\xi^\prime\|
\sum_{i\in \Xi,j\in \Xi^\prime}\big(
\left\|\left[x_i(t),x_{j}\right]\right\|
+\left\|\left[x_i(t),p_{j}\right]\right\| \\
& \hspace{3.5cm}+\left\|\left[p_i(t),x_{j}\right]\right\|+
\left\|\left[p_i(t),p_{j}\right]\right\|
\big)\\
&\le \|\xi\|\|\xi^\prime\|
\sum_{i\in \Xi,j\in \Xi^\prime} f(dist(i,j)),
\end{split}
\end{equation}
where
\begin{equation}
\sum_{i\in \Xi,j\in \Xi^\prime}f(dist(i,j))=\sum_{d=dist(\Xi,\Xi^\prime)}^\infty f(d)\sum_{i\in \Xi,j\in \Xi^\prime}\delta_{dist(i,j),d}.
\end{equation}
We now proceed by showing how to restrict the latter sum to subsets of $\Xi$ and $\Xi^\prime$. As one has to cross the boundary
of a set to find a path to a site outside that set, there exist for all $i\in\Xi$, $j\in\Xi^\prime$ sites $k\in\partial\Xi$, $l\in\partial\Xi^\prime$ 
such that 
\begin{equation}
	dist(i,j)=dist(i,k)+dist(k,l)+dist(l,j).
\end{equation}
Then
$d=dist(i,j)$ requires $dist(i,k)$ and $dist(l,j)$ to be smaller than $d-dist(\Xi,\Xi^\prime)=:r$ as $dist(k,l)\ge dist(\Xi,\Xi^\prime)$. We may thus write
\begin{equation}
\sum_{i\in \Xi,j\in \Xi^\prime}\delta_{dist(i,j),d}=\sum_{i\in \partial\Xi_{r},j\in \partial\Xi^\prime_{r}}\delta_{dist(i,j),d},
\end{equation}
where we denoted by
\begin{equation}
\partial A_r=\bigcup_{i\in\partial A}\left\{j\in A\,\big |\,dist(i,j)\le r\right\}
\end{equation}
the set of lattice sites that are within $A$ and within a layer of thickness $r$ around the surface of $A$, for which we have
\begin{equation}
\begin{split}
\left|\partial A_r\right|&\le\left|\partial A\right|\sup_{i\in \partial A}\left|\left\{j\in A\,\big |\,dist(i,j)\le r\right\}\right|\\
&\le \left|\partial A\right|\sup_{i\in L}\left|\left\{j\in L\,\big |\,dist(i,j)\le r\right\}\right|\\
&= \left|\partial A\right|\sup_{i\in L} \sum_{l=0}^r\left|S_l(i)\right|
\le \left|\partial A\right|\left(1+c_D\sum_{l=1}^rl^{D-1}\right).
\end{split}
\end{equation}
Hence
\begin{equation}
\begin{split}
\sum_{i\in \partial\Xi_r,j\in \partial\Xi^\prime_r}\delta_{dist(i,j),d}
&\le \min\left\{\left|\partial\Xi_{r}\right|,\left|\partial\Xi^\prime_{r}\right|\right\}\sup_{j\in L}S_d(j)\\
&\le c_D\min\left\{\left|\partial\Xi_{r}\right|,\left|\partial\Xi^\prime_{r}\right|\right\}d^{D-1}\\
&\le c_D\min\left\{\left|\partial\Xi\right|,\left|\partial\Xi^\prime\right|\right\}d^{D-1}\left(1+c_D\sum_{l=1}^rl^{D-1}\right)\\
&\le c_D\min\left\{\left|\partial\Xi\right|,\left|\partial\Xi^\prime\right|\right\}d^{D-1}\left(1+c_Dr^D\right).
\end{split}
\end{equation}
To summarize,
\begin{equation}
\begin{split}
\sum_{i\in \Xi,j\in \Xi^\prime}f(dist(i,j))\le&\,
c_D\min\left\{\left|\partial\Xi\right|,\left|\partial\Xi^\prime\right|\right\}\\
&\times
\sum_{d=dist(\Xi,\Xi^\prime)}^\infty f(d)d^{D-1}\left(1+c_D(d-dist(\Xi,\Xi^\prime))^D\right).
\end{split}
\end{equation}
Under the assumptions of Theorem \ref{weyl_appl}, e.g., we may choose 
\begin{equation}
f(dist(i,j))=\left(\frac{\|P\|}{\sqrt{\|PX\|}}+\frac{\|X\|}{\sqrt{\|XP\|}}+2\right)\frac{\me^{d_{i,j}\log\left(\me\tau /d_{i,j}\right)}}{\sqrt{d_{i,j}}\left(1-\left(\me\tau /d_{i,j}\right)^2\right)},
\end{equation}
i.e., for $\me\tau <dist(\Xi,\Xi^\prime)/R=:d_{\Xi,\Xi^\prime}$,
\begin{equation}
\begin{split}
\left\|\left[\hat{W}_{\xi}(t),\hat{W}_{\xi^\prime}\right]\right\|
\le &\frac{c_D \|\xi\|\|\xi^\prime\|\min\left\{\left|\partial\Xi\right|,\left|\partial\Xi^\prime\right|\right\}\left(\frac{\|P\|}{\sqrt{\|PX\|}}+\frac{\|X\|}{\sqrt{\|XP\|}}+2\right)}{\sqrt{d_{\Xi,\Xi^\prime}}\left(1-\left(\me\tau /d_{\Xi,\Xi^\prime}\right)^2\right)}
\me^{d_{\Xi,\Xi^\prime}\log\left(\me\tau /d_{\Xi,\Xi^\prime}\right)}\\
&\times
\sum_{d=0}^\infty 
\me^{d\log\left(\me\tau R/(d+dist(\Xi,\Xi^\prime))\right)/R}
(d+dist(\Xi,\Xi^\prime))^{D-1}\left(1+c_Dd^D\right),
\end{split}
\end{equation}
where we have for the sum the following upper bound ($z:=\me\tau /d_{\Xi,\Xi^\prime}$)
\begin{equation}
\begin{split}
dist^{D-1}(\Xi,\Xi^\prime)
\sum_{d=0}^\infty 
\me^{d\log\left(\me\tau R/(d+dist(\Xi,\Xi^\prime))\right)/R}
(d+1)^{D-1}\left(1+c_Dd^D\right)\\
\le dist^{D-1}(\Xi,\Xi^\prime)
\sum_{d=0}^\infty z^{d/R}
(d+1)^{D-1}\left(1+c_D(d+1)^D\right).
\end{split}
\end{equation}

\section{Summary}

In this work, we have presented Lieb-Robinson bounds for harmonic
lattice systems on general lattices, 
complementing and generalizing work in 
Refs.\ \cite{Schlein} (see also Ref.\ \cite{Oliver,Oliver2}). 
We found a stronger than exponential decay in case of local
interactions, and an inheritance of the decay behavior in case of 
algebraically decaying interactions. For the case of the Klein-Gordon
field, we found the exact locality emerging from the approximate
locality in the Lieb-Robinson sense. Specific attention was devoted
to the time evolution of Weyl operators, which are an 
important class of operators in harmonic lattices. 
As such, this work provides a framework to study 
non-equilibrium dynamics in harmonic lattice
systems in a general setting.

\section{Acknowledgements}

This work has been supported by
the EU (QAP), the EPSRC, Microsoft Research,
and the European Research Councils (EURYI).
Note that in independent work, Refs.\ 
\cite{Oliver,Oliver2} came to similar conclusions as the ones 
presented in this work.

\end{document}